\documentclass[lettersize,journal]{IEEEtran}
\usepackage{tikz}
\usepackage{booktabs}
\usepackage{amsmath,amsfonts}
\usepackage{algorithmic}
\usepackage{algorithm}
\usepackage{array}
\usepackage{textcomp}
\usepackage{stfloats}
\usepackage{url}
\usepackage{verbatim}
\usepackage{graphicx}
\usepackage{cite}
\usepackage{subfigure}
\usepackage{color,soul}
\usepackage{bm}
\usepackage{epstopdf}
\usepackage{extarrows}
\makeatletter

\newcommand{\Rmnum}[1]{\expandafter\@slowromancap\romannumeral #1@}
\makeatother
\usepackage{xcolor}
\usepackage{xpatch}

\begin{document}
	\title{Pinching-Antenna-Enabled Cognitive Radio Networks}
	
	\author{Zeyang~Sun,~\IEEEmembership{Graduate~Student~Member,~IEEE},
		Xidong~Mu,~\IEEEmembership{Member,~IEEE},
		Shuai~Han,~\IEEEmembership{Senior~Member,~IEEE},
		Sai~Xu,~\IEEEmembership{Member,~IEEE},	
		and~Michail~Matthaiou,~\IEEEmembership{Fellow,~IEEE}%
		\thanks{This work was supported in part by the National Key Research and Development Program of China under Grant 62371166, in part by the National Natural Science Foundation of China under Grant 624B2052. Part of this paper has been submitted to the 2026 IEEE ICC \cite{Zeyang4}. (\textit{{Corresponding author: Shuai Han.}})}
		\thanks{Z. Sun and S. Han are with the School of Electronics and Information Engineering, Harbin Institute of Technology, Harbin 150001, China (e-mail: zeyangsun97@gmail.com; hanshuai@hit.edu.cn).}
		\thanks{X. Mu and M. Matthaiou are with the Centre for Wireless Innovation, Queen's University Belfast, Belfast BT3 9DT, U.K. (e-mail: x.mu@qub.ac.uk; m.matthaiou@qub.ac.uk).}
		\thanks{S. Xu is with the Department of Electronic and Electrical Engineering, University College London, London WC1E 6BT, U.K. (e-mail: sai.xu@ucl.ac.uk).}
		}
	\maketitle 
	\begin{abstract}
		This paper investigates a pinching-antenna (PA)-enabled cognitive radio network, where both the primary transmitter (PT) and secondary transmitter (ST) are equipped with a single waveguide and multiple PAs to facilitate simultaneous spectrum sharing. Under a general Ricean fading channel model, a closed-form analytical expression for the average spectral efficiency (SE) achieved by PAs is first derived. Based on this, a sum-SE maximization problem is formulated to jointly optimize the primary and secondary pinching beamforming, subject to system constraints on the transmission power budgets, minimum antenna separation requirements, and feasible PA deployment regions. To address this non-convex problem, a three-stage optimization algorithm is developed to sequentially optimize both the PT and ST pinching beamforming, and the ST power control. For the PT and ST pinching beamforming optimization, the coarse positions of PA are first determined at the waveguide-level. Then, wavelength-level refinements achieve constructive signal combination at the intended user and destructive superposition at the unintended user. For the ST power control, a closed-form solution is derived. Simulation results demonstrate that i) PAs can achieve significant SE improvements over conventional fixed-position antennas; ii) the proposed pinching beamforming design achieves effective interference suppression and superior performance for both even and odd numbers of PAs; and iii) the developed three-stage optimization algorithm enables nearly orthogonal transmission between the primary and secondary networks.
	\end{abstract}
	
	\section{Introduction}
	The rapid proliferation of wireless communication devices and their expanding applications across diverse scenarios, ranging from dense urban deployments to emerging Internet-of-Things (IoT) networks, have significantly intensified the demand for spectral resources \cite{IoT1,NOMA_survey_1,Zeyang1}. The scarcity of available spectrum has consequently emerged as a critical bottleneck, challenging the sustained growth of wireless communication systems. To this end, spectrum sharing techniques, particularly cognitive radio (CR), have attracted significant attention by enabling dynamic spectrum access and opportunistic utilization of frequency bands to maximize the spectral efficiency (SE). \cite{CC-CR1,Zeyang2}. CR networks allow a secondary user (SU) to access underutilized frequency bands without adversely impacting the primary user (PU). However, the coexistence of primary and secondary networks inherently introduces severe co-channel interference (CCI), substantially deteriorating the quality and reliability of wireless transmissions \cite{CR0}. Consequently, the exploration of advanced transceiver architectures has emerged as a pivotal challenge in improving both the capacity and reliability of next-generation wireless networks \cite{CR2}.
	
	Multiple-antenna techniques have been widely recognized as a promising solution for CR networks, as they can significantly improve the SE and mitigate CCI through spatial multiplexing and beamforming \cite{CR_WSR_power_constraint,CC-BF1,XuSai1,XuSai2}. Nevertheless, since antennas are deployed at fixed positions, conventional multi-antenna techniques cannot fully avail of the continuous spatial variation of wireless channels, leaving part of the spatial degrees of freedom (DoF) underutilized \cite{MA1,JinmingMA}. To introduce spatial adaptability, flexible-antenna technologies, such as fluid antennas (FAs) \cite{Fluid} and movable antennas (MAs) \cite{MA1,JinmingMA}, have been proposed, enabling wavelength-scale repositioning of the radiating elements. While such designs exploit local spatial diversity to assist interference management, their displacement range remains inherently limited, rendering them vulnerable to line-of-sight (LoS) blockage and incapable of mitigating large-scale path loss \cite{PA1}. These limitations motivate transceiver architectures that provide large-scale spatial reconfigurability, robust establishment of LoS links, and adaptive provisioning of radiating elements within feasible deployment regions.	
	
	To overcome these limitations, the pinching-antenna (PA) technology has recently emerged as a promising solution for spatially reconfigurable wireless transceivers. The PA technology was first proposed by NTT DOCOMO in 2022 \cite{NTT}, which enables dynamic formation of radiating elements by selectively activating specific points along a waveguide using small dielectric particles \cite{PA1,PA2}, thereby realizing leaky-wave antennas with programmable and distributed radiation sites. This architecture supports low-cost and flexible deployment, enabling wide-range spatial reconfiguration without additional hardware, and significantly expands the spatial DoF available for adaptive transceiver design \cite{PA2,Xu2025c,Xu2025f}. 
	
	\subsection{Related Works}
	Given these attractive features, the PA technology has become an emerging research topic, with growing interest in both analytical performance evaluation and algorithm design. The authors of \cite{Ding2025b} analyzed PA systems under single- and multi-waveguide settings, showing that PAs can establish strong LoS links and mitigate path loss compared with conventional antenna arrays. This foundational study motivated subsequent research dedicated to enhancing communication performance through optimized PA pinching beamforming. For example, the authors of \cite{Wang2025a} addressed the transmit power minimization problem by proposing a penalty-based alternating optimization algorithm that jointly optimizes the transmit and pinching beamforming. The authors of \cite{PA-WSR} developed a fractional programming block coordinate descent algorithm to maximize the weighted sum rate in PA-assisted downlink multiple-input multiple-output (MIMO) systems. The authors of \cite{Lv2025} studied the beam training design problem for PA systems in single-waveguide single-user, single-waveguide multi-user, and multi-waveguide multi-user configurations and proposed an efficient three-stage beam training scheme to reduce the training overhead.
		
	Building on these foundational studies, recent research has explored PA technology in broader wireless paradigms, including integrated sensing and communications (ISAC) \cite{PA-ISAC1}, non-orthogonal multiple access (NOMA) \cite{Zhou2025f}, and physical-layer security (PLS) \cite{PA-PLS1}. Specifically, \cite{PA-PLS1} developed a PLS framework for PA systems by jointly optimizing baseband and pinching beamforming to improve the secrecy rate and weighted secrecy sum-rate in both single-user and multiuser scenarios. The authors of \cite{Zhou2025f} investigated a PA-assisted downlink NOMA network, where the sum-rate is maximized by jointly optimizing power allocation and PA placement based on the Karush-Kuhn-Tucker (KKT) conditions and the bisection-based search algorithm. The authors of \cite{PA-ISAC1} investigated a PA-enabled ISAC network and developed a penalty-based alternating optimization algorithm to maximize the illumination power under quality of service (QoS) constraints.
	
	\subsection{Motivations and Contributions}
	Existing works have demonstrated that by exploiting pinching beamforming, the desired signal can be constructively enhanced while harmful interference can be suppressed \cite{PA1,PA2}. These attributes render PA-enabled CR networks particularly suitable for spectrum sharing and interference management. To the best of our knowledge, no prior work has investigated the integration of PAs into CR networks. Moreover, such integration introduces unique technical challenges that remain unexplored in the existing PA literature. First, unlike existing PA systems that optimize pinching beamforming at a single transmitter, PA-enabled CR networks require joint optimization of pinching beamforming at both the primary transmitter (PT) and the secondary transmitter (ST). This joint optimization is inherently challenging due to the interference coupling between their signals. Second, the existing PA studies focus exclusively on pure LoS channels. In PA-enabled CR networks, the large number of radiating elements and links makes pure LoS propagation across all paths unlikely. Reflections, scattering, and diffraction will inevitably introduce non-line-of-sight (NLoS) components. Therefore, neglecting NLoS propagation results in overly idealized channel models, limiting the generalizability of existing PA system studies by failing to incorporate a general channel model that accounts for both LoS and NLoS effects.
	
	Against the above background, we investigate a PA-enabled CR network under a general Ricean fading model. We derive a closed-form average SE (ASE) and study a sum-SE maximization that jointly optimizes the pinching beamforming in the primary network (PN) and secondary network (SN) together with the ST power. Our main contributions are summarized as follows:

	\begin{itemize}
		\item We propose a PA-enabled CR framework, wherein both the PT and ST are equipped with a single waveguide and multiple PAs to enable spectrum sharing. We consider a general Ricean fading channel model that captures both LoS and NLoS components in the PA-enabled CR network. Based on this, we derive a closed-form expression for the ASE based on the statistical properties of Ricean fading channels. Subsequently, we formulate a sum-SE maximization problem, subject to constraints on the PA deployment region, minimum inter-antenna separation, and the ST’s power budget.
		\item We develop an efficient three-stage optimization algorithm to successively optimize the placement of the primary and secondary PAs, and the power control of the ST. For the pinching beamforming at both PT and ST, we propose a hierarchical optimization framework for the PA location that jointly accounts for large-scale propagation and fine-grained phase alignment. At the waveguide level, spatial placement is optimized to reduce path loss and strengthen link quality. At the wavelength level, wavelength-scale phase control enforces constructive combining at the intended receiver and destructive combining at the unintended ones. In particular, for even-numbered and odd-numbered PAs, we respectively develop the new fixed phase assignment and refined phase assignment schemes to further enhance interference suppression capability.
		\item Extensive simulations indicate that 1) the proposed PA-enabled CR network attains measurable improvements over conventional fixed-position antennas; 2) the proposed algorithm exhibits robust performance across both the even- and odd-numbered PAs; and 3) with the deployment of PAs, the PN and SN achieve nearly orthogonal transmission even when the PT and ST are close to each other.
	\end{itemize}
	
	\subsection{Organization and Notations}
	The remainder of this paper is organized as follows: Section II introduces the system model of the PA-enabled CR network, including the channel assumptions and key performance metrics, and formulates the sum-SE maximization problem. Section III develops a three-stage optimization algorithm for joint pinching beamforming and power allocation. Section IV provides simulation results to demonstrate the performance advantages of the proposed approach under various network scenarios. Finally, Section V concludes the paper.
	
	\textit{Notations}: Scalars, vectors and matrices are denoted by lowercase, boldface lowercase, and boldface uppercase letters, respectively; $|x|$ denotes the modulus of complex scalar $x$; $\|\mathbf{x}\|$ denotes the Euclidean norm of vector $\mathbf{x}$; ${\rm{Tr}}\left( {\bf{X}} \right)$ denotes the trace of matrix $\bf X$. The superscripts ${\left(  \right)^*}$ and ${\left(  \right)^T}$ denote the conjugate and transpose, respectively; $\mathbb{R}$ and $\mathbb{Z}^{+}$ denote the sets of real numbers and positive integers, respectively; ${\mathbb E}\left\{\!{\cdot}\!\right\}$ represents the expectation operator. Finally, $\mathcal{CN}(\mu,\sigma ^2)$ denotes a circularly symmetric complex Gaussian distribution with mean $\mu$ and variance $\sigma ^2$. 
	
	\section{System Model and Problem Formulation}
	In this section, we introduce the system model for the PA-enabled CR network, derive the ASE under the general Ricean channel model, and formulate the joint pinching beamforming and ST power allocation optimization problem to maximize the sum ASEs.
	\subsection{System Model}\label{sec2-1}
	As illustrated in Fig. 1, we investigate a PA-enabled downlink CR network, which consists of a PN and SN. The PN comprises a PT employing a single waveguide and activates $N$ PAs to serve a single-antenna PU $u_p$. Similarly, the ST employs a single dielectric waveguide, along which $M$ PAs are strategically deployed to simultaneously communicate with a single-antenna SU $u_s$ under the same frequency band of the PN. All PAs are capable of dynamic repositioning along the waveguide. Without loss of generality, we assume that the lengths of all waveguides are equal to $L_x$, and they are deployed parallel to the $x$-axis and at an altitude of $l$.
	
	\begin{figure}[tpb]
		\centering
		\includegraphics[width=3.5in]{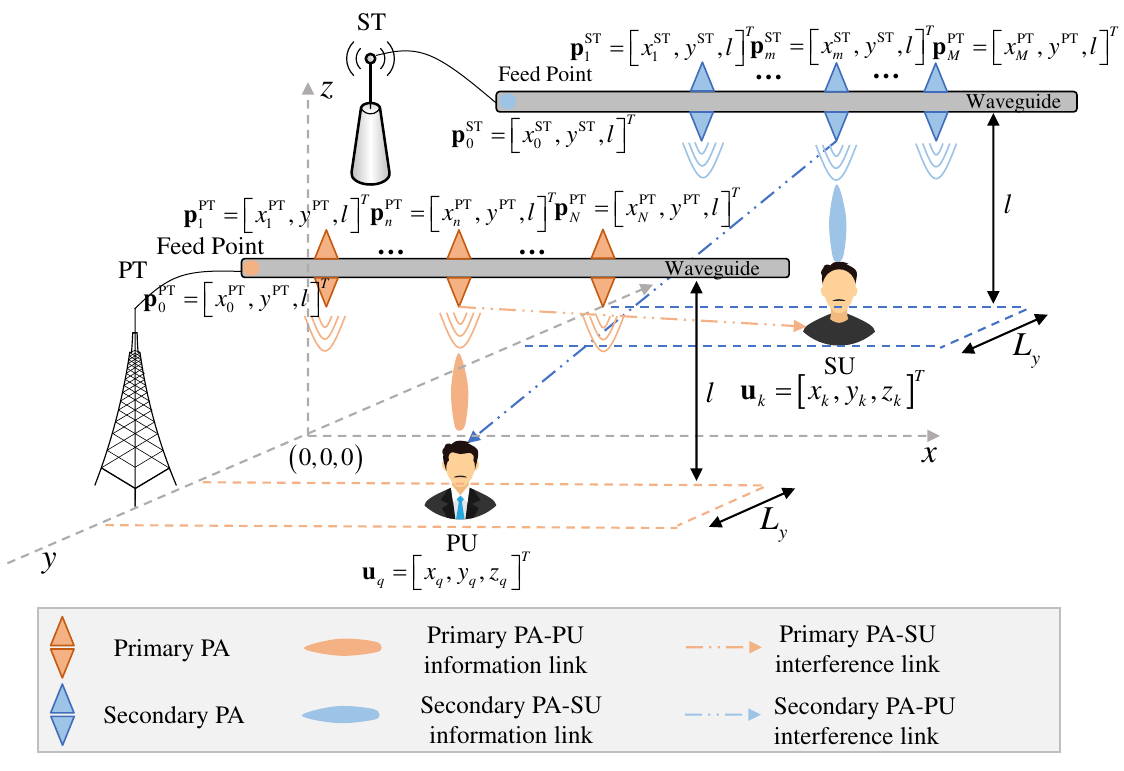}
		\caption{An illustration of a PA-enabled CR network.}
		\label{Fig1}
	\end{figure}

	We adopt a 3D Cartesian coordinate system with both waveguides parallel to the $x$-axis and all PAs lying on the plane $z=l$. Let $\mathcal I=\{\mathrm{PT},\mathrm{ST}\}$ and $\mathcal R=\{p,s\}$ denote the transmitter set and the receiver set, respectively. We use a unified receiver index $r\in\mathcal{R}$ to denote the PU ($r=p$) and SU ($r=s$), whose locations are $\mathbf{u}_r=\big[x_r,\,y_r,\,z_r\big]^T,\ r\in\{p,s\}$. A transmitter identity is indexed by $i\in\mathcal{I}$, and its waveguide feed point is $\mathbf{p}_0^{\,i}=\big[\,x_0^{\,i},\,y^{\,i},\,l\,\big]^T$,
	where $x_0^{\,i}$ is the $x$-coordinate of the feed point and $y^{\,i}$ is the $y$-coordinate of that waveguide. For transmitter $i$, denote its PA index set by $\mathcal N_i$: $\mathcal N_{\mathrm{PT}}=\mathcal N=\{1,\ldots,N\}$ and $\mathcal N_{\mathrm{ST}}=\mathcal M=\{1,\ldots,M\}$. A PA attached to transmitter $i$ is indexed by $t\in\mathcal N_i$ and is located at $\mathbf p_t^{\,i}=[x_t^{\,i},\,y^{\,i},\,l]^T$. For notational convenience, we write $t=n,$ when $t\in\mathcal N,i=\mathrm{PT}$ and $t=m$ when $i=\mathrm{ST},t\in\mathcal M$. Furthermore, the $x$-coordinates of transmitter $i$'s PAs is collected as $\mathbf x^{\,i}=[x_1^{\,i},\ldots,x_{|\mathcal N_i|}^{\,i}]^T$, where $|{{{\mathcal N}_i}}|$ denotes the number of PAs for $i \in \mathcal R$.
	
	The in-waveguide channel from the transmitter feed point to user $r$ through PA $t$ can be written as  
	\begin{equation}
			{g_t} = \begin{cases}
			{{e^{ - j\frac{{2\pi }}{{{\lambda _g}}}\left\| {{\bf{p}}_0^{{\rm{PT}}} - {\bf{p}}_n^{{\rm{PT}}}} \right\|}}}, \ t \in {\cal N}, \\
			{{e^{ - j\frac{{2\pi }}{{{\lambda _g}}}\left\| {{\bf{p}}_0^{{\rm{ST}}} - {\bf{p}}_m^{{\rm{ST}}}} \right\|}}}, \ t \in {\cal M},
		\end{cases}
	\end{equation}
	where ${g_{t}}$ denotes the in-waveguide phase shift accounting for propagation from feed to PA $t$, while ${h_{t,r}}$ is the channel coefficient from the $t$-PA to the user $r$. The guided wavelength is ${\lambda _g} = \frac{\lambda }{{{\eta _{{\rm{eff}}}}}}$ \cite{PA-carefully1,Qian2025}, with $\lambda$ and ${{\eta _{{\rm{eff}}}}}$ representing the free-space wavelength and the effective refractive index of the dielectric waveguide, respectively. 	

    For the PA-user channel, most existing studies on PA systems consider only LoS links between PAs and users (e.g., $h_{t,r}$) \cite{Xu2025c,Xu2025f,PA-ISAC1,PA-PLS1,PA-PLS2}, neglecting the NLoS components arising from reflections, scattering, and multipath effects. In practical wireless environments, such as urban and indoor scenarios, NLoS propagation significantly impacts channel characteristics and link reliability. Consequently, we assume that the relevant PA-to-user channels follow the Ricean channel model, which can be formulated as 
	\begin{equation}
		\begin{aligned}
			{h_{t,r}} = \sqrt {\frac{\eta }{{(\kappa  + 1){{\left\| {{{\bf{u}}_r} - {{\bf{p}}_t}} \right\|}^{\chi}}}}}& \left( {\sqrt \kappa  {{\bar h}_{t,r}} + {{\tilde h}_{t,r}}} \right), t \in \mathcal{N}_i, \ r \in \mathcal R, \\
		\end{aligned}
		\label{eq1}
	\end{equation}
	where ${\kappa}$ and $\chi >2$ denote the Ricean factor and path-loss exponent, respectively; $\eta  = \frac{{{c^2}}}{{16{\pi ^2}f_c^2}}$ represents the channel gain at the reference distance, which is determined by the carrier frequency $f_c$ and the speed of light $c$. Furthermore, $\bar{h}_{t,r} = 
	\begin{cases}
		e^{-j\frac{2\pi}{\lambda}\| \mathbf{u}_r - \mathbf{p}_n^{\mathrm{PT}}\|}
		& t \in \mathcal{N}, \\[15pt]
		e^{-j\frac{2\pi}{\lambda}\| \mathbf{u}_r - \mathbf{p}_m^{\mathrm{ST}}\|}
		& t \in \mathcal{M},
	\end{cases}$ and $\tilde{h}_{t,r}$ denote the deterministic LoS component and the corresponding NLoS Rayleigh fading component for the $t$-th PA to user $u_r$ link, respectively. 
	
	An equal power allocation model is considered wherein the transmit power is uniformly distributed across multiple PAs \cite{Wang2025a,Xabc}. Based on the established system framework, the received signal at the PU is given by
	\begin{equation}
		\begin{aligned}
			{y_p} = \sqrt {\frac{{{P_{{\rm{PT}}}}}}{N}} \sum\limits_{n = 1}^N {{h_{n,p}}{g_n}x}  + \sqrt {\frac{{{p_{\rm ST}}}}{M}} \sum\limits_{m = 1}^M {{h_{m,p}}{g_m}c}  + {z_p},
		\end{aligned}
		\label{eq3}
	\end{equation}
	where ${P_i}$ and $p_{\rm ST} \leq P_{\mathrm{ST}}$ represent the predefined peak power budget for $i \in \mathcal I$ and the transmit power at the ST,{\footnote{In practical CR networks, the ST must dynamically adjust its transmit power $p_{\rm ST}$ to avoid the excessively powerful interference to PU, while the PT operates at its maximum allowable power ${P_{\mathrm{PT}}}$ to guarantee QoS of PU. }} respectively; $x$ and $c$ denote the signal intended for PU and SU with ${\mathbb E} \left\{ {{{\left| x \right|}^2}} \right\} = 1$ and ${\mathbb E} \left\{ {{{\left| c \right|}^2}} \right\} = 1$, respectively; $z_p \sim {\cal C}{\cal N}\left( {{{0}},\sigma^2} \right)$ is the additive white Gaussian noise (AWGN) at the PU $u_p$.

	Consequently, the signal-to-interference-noise ratio (SINR) of the PU is given by
	\begin{equation}
		\begin{aligned}
			{\gamma _p} = \frac{{\frac{{{P_{{\rm{PT}}}}}}{N}{{\left| {\sum\limits_{n = 1}^N {{h_{n,p}}{g_n}} } \right|}^2}}}{{\frac{{{p_{\rm ST}}}}{M}{{\left| {\sum\limits_{m = 1}^M {{h_{m,p}}{g_m}} } \right|}^2} + {\sigma ^2}}},
		\end{aligned}
		\label{eq3_1}
	\end{equation}
	while the corresponding achievable SE for PU is written as $R_p = {{{\log }_2}\left( {1 + \gamma _p} \right)}$.
	
	Due to the spectrum sharing between the PN and SN, the SN will entail CCI to the PU. Therefore, the interference temperature constraint (ITC) is employed to ensure the communication performance of the PU, which can be modeled as
	\begin{equation}
		\frac{{{p_{\rm ST}}}}{M}{\left| {\sum\limits_{m = 1}^M {{h_{m,p}}{g_m}} } \right|^2} \le {P^{{\rm{TH}}}},
		\label{eq4}
	\end{equation}
	where ${P^{{\rm{TH}}}}$ is the predefined maximum interference power.
	
	Similarly, the signal received by the SU $u_s$ is denoted as
	\begin{equation}
		\begin{aligned}
			{y_s} = \sqrt {\frac{{{p_{\rm ST}}}}{M}} \sum\limits_{m = 1}^M {{h_{m,s}}{g_m}c}  + \sqrt {\frac{{{P_{{\rm{PT}}}}}}{N}} \sum\limits_{n = 1}^N {{h_{n,s}}{g_n}x}  + {z_s},
		\end{aligned}
		\label{eq5}
	\end{equation}
	where $z_s \sim {\cal C}{\cal N}\left( {{{0}},\sigma^2} \right)$ is the AWGN at the SU $u_s$.
	
	The SINR for the SU to decode its own signal is given by
	\begin{equation}
		\begin{aligned}
			{\gamma _s} = \frac{{\frac{{{p_{\rm ST}}}}{M}{{\left| {\sum\limits_{m = 1}^M {{h_{m,s}}{g_m}} } \right|}^2}}}{{\frac{{{P_{{\rm{PT}}}}}}{N}{{\left| {\sum\limits_{n = 1}^N {{h_{n,s}}{g_n}} } \right|}^2} + {\sigma ^2}}},
		\end{aligned}
		\label{eq6}
	\end{equation}
	where the corresponding achievable SE for SU is denoted as $R_s = {{{\log }_2}\left( {1 + \gamma _s} \right)}$.
	
	\subsection{Performance Metric}
	The channel coefficients ${h_{t,r}}$ are inherently stochastic due to the presence of Rayleigh distributed NLoS components. Consequently, the SINR of the PU and SU, as well as the interference power at the PU, are characterized as random variables, making it infeasible to obtain deterministic expressions for instantaneous achievable SEs. To effectively evaluate the system performance, we adopt the average achievable SE \cite{NLoS}, defined as ${\bar R_p} = \mathbb E \left\{ {{R_p}} \right\}$ and ${\bar R_s} = \mathbb E \left\{ {{R_s}} \right\}$, as a performance metric for the PU and SU. Similarly, the instantaneous ITC is reformulated as $\frac{{{p_{{\rm{ST}}}}}}{M}E\left\{ {{{\left| {\sum\limits_{m = 1}^M {{h_{m,p}}{g_m}} } \right|}^2}} \right\} \le {P^{{\rm{TH}}}}$ as a mathematically tractable surrogate, while still ensuring reliable PU protection. However, the derivation of a closed-form expression for the expectation terms poses significant mathematical challenges due to the underlying stochastic channel distributions. To overcome this limitation, we leverage \textit{Lemma} 1 and \textit{Lemma} 2 to establish an approximate analytical expression for the expected achievable SEs for users. 
	
	\textit{Lemma 1}: For arbitrary positive constant $a,b$ and independent non-negative random variables $X,Y$, the following asymptotic equivalence holds
	\begin{equation}
		\begin{aligned}
			\mathbb E \left\{ {{{\log }_2}\left[ {1 + \frac{a}{{b + \frac{X}{Y}}}} \right]} \right\} \approx   {{{\log }_2}\left[ {1 + \frac{a}{{b + \frac{{\mathbb E {\left\{ X \right\}}}}{\mathbb E{{\left\{ Y \right\}}}}}}} \right]}.
		\end{aligned}
		\label{eq7}
	\end{equation}
	
	\ \ Proof: The proof is omitted for brevity, as it closely resembles that of [Theorem 1, 32].
		
	\textit{Lemma 2}: The expected values for the channel gains from the $t$-th PA to the user $r$ can be derived as follows 
	\begin{equation}
		\begin{aligned}
			\mathbb E\left\{ {{{\left| {\sum\limits_t {{h_{t,r}}{g_t}} } \right|}^2}} \right\} \buildrel \Delta \over = \psi _r^i, \ i \in {\cal I},
		\end{aligned}
		\label{eq7_1}
	\end{equation}
	where for $r \in \{ p, s \}$, the expressions of $\psi _r^{{\rm{PT}}}$ and $\psi _r^{{\rm{ST}}}$ are expressed as
	\begin{small}
		\begin{equation}
			\begin{aligned}
				\psi _r^{{\rm{PT}}} = \frac{\eta }{{\kappa  + 1}}\left( {\kappa {{\left| {\sum\limits_{n \in \mathcal N} {\frac{{{{\bar h}_{n,r}{g_n}}}}{{\left\| {{{\bf{u}}_r} - {{\bf{p}}_n^{{\rm{PT}}}}} \right\|}^{\frac{\chi }{2}}}} } \right|}^2} + \sum\limits_{n \in \mathcal N} {\frac{1}{{{{\left\| {{{\bf{u}}_r} - {{\bf{p}}_n^{{\rm{PT}}}}} \right\|}^{\chi}}}}} } \right),
			\end{aligned}
			\label{eq7_2}
		\end{equation}
		\begin{equation}
			\begin{aligned}
				\psi _r^{{\rm{ST}}} = \frac{\eta }{{\kappa  + 1}}\left( {\kappa {{\left| {\sum\limits_{m \in \mathcal M} {\frac{{{{\bar h}_{m,r}{g_m}}}}{{\left\| {{{\bf{u}}_r} - {{\bf{p}}_m^{{\rm{ST}}}}} \right\|}^{\frac{\chi }{2}}}} } \right|}^2} + \sum\limits_{m \in \mathcal M} {\frac{1}{{{{\left\| {{{\bf{u}}_r} - {{\bf{p}}_m^{{\rm{ST}}}}} \right\|}^{\chi}}}}} } \right).
			\end{aligned}
			\label{eq7_3}
		\end{equation}
	\end{small}
	\ \ Proof: Appendix~A.
	
	Based on \textit{Lemma 1} and \textit{Lemma 2}, the expectation expression for the ASE of PU and SU can be approximately written as
	\begin{equation}
		\begin{aligned}
			{{\bar R}_p} \approx {\log _2}\left[ {1 + \frac{{M{P_{{\rm{PT}}}}\psi _p^{{\rm{PT}}}}}{{N{p_{\rm ST}}\psi _p^{{\rm{ST}}} + NM{\sigma ^2}}}} \right],
		\end{aligned}
		\label{eq10_0}
	\end{equation}
	\begin{equation}
		\begin{aligned}
			{{\bar R}_s} \approx  {\log _2}\left[ {1 + \frac{{N{p_{\rm ST}}\psi _s^{{\rm{ST}}}}}{{M{P_{{\rm{PT}}}}\psi _s^{{\rm{PT}}} + NM{\sigma ^2}}}} \right].
		\end{aligned}
		\label{eq10}
	\end{equation}

	\subsection{Sum of Averaged Achievable SE Maximization Problem}
	In this paper, we aim to maximize the sum of the ASEs of both the PU and the SU, denoted as ${{\bar R}^{{\rm{Sum}}}} = {{\bar R}_p} + {{\bar R}_s}$. This metric provides a comprehensive assessment of the system's capacity by jointly considering the performance of both the PU and SU. Unlike conventional frameworks that focus solely on maximizing the SU's SE subject to the ITC of the PU, our formulation seeks to enhance the overall SEs of the proposed CR network, thus providing a system-level optimization objective. Given the hierarchical nature of CR networks, where the PU retains transmission priority, it is crucial to ensure the PU's QoS at all times. Accordingly, we enforce an ITC that limits the aggregate interference caused by the secondary PAs at the PU below a predefined threshold. Therefore, the sum-SE maximization problem can be mathematically formulated as
	\begin{subequations}
		\begin{equation}
			\begin{aligned}
				({\rm{P1}})\mathop {\max }\limits_{{p_{\rm ST}},{{\bf{x}}^{{\rm{PT}}}},{{\bf{x}}^{{\rm{ST}}}}} {{\bar R}^{{\rm{Sum}}}} 
			\end{aligned}
			\label{eq14_1}
		\end{equation}
		\begin{equation}
			\begin{aligned}
				{\mathrm{ s.t.}} \ 0 < {p_{\rm ST}} \le P_{\rm ST},
			\end{aligned}
			\label{eq14_2}
		\end{equation}
		\begin{equation}
			\begin{aligned}
				\frac{{{p_{\rm ST}}}}{M} \psi _p^{{\rm{ST}}} \le {P^{{\rm{TH}}}},
			\end{aligned}
			\label{eq14_3}
		\end{equation}
		\begin{equation}
			\begin{aligned}
				x_{t}^{{{i}}} - x_{t-1}^{{{i}}} \ge \Delta {_{{\rm{min}}}},
			\end{aligned}
			\label{eq14_4}
		\end{equation}
		\begin{equation}
			\begin{aligned}
				x_t^{{{i}}} \in {{\cal S}^{{\rm{i}}}},
			\end{aligned}
			\label{eq14_6}
		\end{equation}
	\end{subequations}
	where (\ref{eq14_2}) restricts the transmission power of ST, (\ref{eq14_3}) enforces the ITC at the PU, (\ref{eq14_4}) regulates the minimum space $\Delta {_{{\rm{min}}}}$ for PAs to avoid their mutual coupling, while (\ref{eq14_6}) specifies the feasible positioning domains for the PAs.
	
	The optimization problem (\rm P1) is analytically and computationally challenging. The primary challenge arises from the strong coupling between the positions of the primary and secondary PAs in determining both users' ASEs. Specifically, the configuration of the primary PAs simultaneously affects not only the signal quality at the PU, but also the interference experienced by the SU, and vice versa for the secondary PAs. This interdependence results in a highly non-convex optimization problem, where the objective function is a complex function of the joint pinching beamformer design. In addition, the optimization problem is further complicated by the existence of non-convex constraints, including the requirements for minimum antenna spacing and adherence to interference temperature limits. These features make the problem not amenable to standard convex optimization techniques, rendering the search for a global optimum intractable in general. Consequently, efficient alternative algorithmic strategies or appropriate approximations must be devised to jointly optimize the antenna locations and power allocation. To tackle these challenges, we will develop a three-stage optimization algorithm.

	\section{Proposed Three-Stage Optimization Algorithm}
    \label{Proposed}
    In this section, we first reformulate the objective function to provide analytical insights that guide the solution of problem (\rm P1). Based on this, we introduce a three-stage optimization algorithm that implements the waveguide-level and wavelength-level pinching beamforming optimizations, followed by the ST power control.    

	\subsection{Objective Function Reformulation}
	Substituting \eqref{eq7_2} and \eqref{eq7_3} into \eqref{eq14_1}, the objective function can be reformulated as
	\begin{equation}
		\begin{aligned}
		{{\bar R}^{{\rm{Sum}}}}& = {{\bar R}_p} + {{\bar R}_s}\\
		&= {\log _2}\left[ {1 + \frac{{M{P_{{\rm{PT}}}}\psi _p^{{\rm{PT}}}}}{{Np_{{\rm{ST}}}\psi _p^{{\rm{ST}}} + NM{\sigma ^2}}}} \right]\;\\
		&+ {\log _2}\left[ {1 + \frac{{Np_{{\rm{ST}}}\psi _s^{{\rm{ST}}}}}{{M{P_{{\rm{PT}}}}\psi _s^{{\rm{PT}}} + NM{\sigma ^2}}}} \right].
		\end{aligned}
	\end{equation}
	
	Note that for the optimal ST power ${p_{{\rm{ST}}}^{*}}$, the objective function maximization problem is equal to
	\begin{equation}
		\begin{aligned}
			\max & \ {{\bar R}^{{\rm{Sum}}}}\\
			\Leftrightarrow \max \left( {\psi _p^{{\rm{PT}}},\psi _s^{{\rm{ST}}}} \right) \ &{\rm{ and}} \ \min \left( {\psi _p^{{\rm{ST}}},\psi _s^{{\rm{PT}}}} \right),
		\end{aligned}
	\end{equation}
	where $\max \left( {\psi _p^{{\rm{PT}}},\psi _s^{{\rm{ST}}}} \right)$ corresponds to the case with optimal PT and ST pinching beamformers that maximize the desired signal for the intended user, while $\min \left( {\psi _p^{{\rm{ST}}},\psi _s^{{\rm{PT}}}} \right)$ suppresses the interference for unintended user. The joint maximization of ${\psi_p^{\rm{PT}}}$ and ${\psi_s^{\rm{ST}}}$, along with the minimization of ${\psi_p^{\rm{ST}}}$ and ${\psi_s^{\rm{PT}}}$, quantifies the optimal primary and secondary PA positions that simultaneously enhance constructive beamforming gains and suppress mutual interference between the PN and SN.
%
	
	The reformulated objective function remains challenging due to the coupling between the optimization variables ${\bf x}^{\rm PT}$ and ${\bf x}^{\rm ST}$. Nevertheless, by inspecting the objective function's structure, we observe that the primary PA positions $\mathbf{x}^{\text{PT}}$ affect the objective solely through the terms $\psi_p^{\text{PT}}$ and $\psi_s^{\text{PT}}$, while the secondary PA positions $\mathbf{x}^{\text{ST}}$ affect it solely through $\psi_p^{\text{ST}}$ and $\psi_s^{\text{ST}}$. This coupling exhibits a block-separable structure, which motivates a three-stage optimization algorithm that decomposes the original problem into sequential steps for primary pinching beamforming, secondary pinching beamforming, and ST power control, as detailed subsequently.
	
	\subsection{Primary Pinching Beamforming Optimization Design}
	\label{PTPA_OP}
	\subsubsection{Problem Reformulation}
	\label{PTPA_OP1}
	With the given $\left( {{p_{\rm ST}},{{\bf{x}}^{{\rm{ST}}}}} \right)$, the objective function can be rewritten as
	\begin{equation}
		\mathop {\max }\limits_{{{\bf{x}}^{{\rm{PT}}}}} \psi _p^{{\rm{PT}}}\left( {{{\bf{x}}^{{\rm{PT}}}}} \right) \ {\rm{ and }} \ \mathop {\min }\limits_{{{\bf{x}}^{{\rm{PT}}}}} \psi _s^{{\rm{PT}}}\left( {{{\bf{x}}^{{\rm{PT}}}}} \right).
	\label{P2}
	\end{equation}
	
	Substituting (\ref{eq7_2}) and (\ref{eq7_3}) into (\ref{P2}), then the sum SE maximization problem (\rm P1) can be transformed into 
	\begin{subequations}
		\begin{equation}
			\begin{aligned}
				\left( {{\rm{P2}}} \right)\;\mathop {\max }\limits_{x_1^{{\rm{PT}}},...,x_N^{{\rm{PT}}}} \kappa {\left| {\sum\limits_{n \in {\cal N}} {\frac{{{e^{ - j{\phi _{n,p}}}}}}{{{{\left\| {{{\bf{u}}_p} - {\bf{p}}_n^{{\rm{PT}}}} \right\|}^{\frac{\chi }{2}}}}}} } \right|^2} + \sum\limits_{n \in {\cal N}} {\frac{1}{{{{\left\| {{{\bf{u}}_p} - {\bf{p}}_n^{{\rm{PT}}}} \right\|}^\chi }}}} \\
				\;\mathop {\min }\limits_{x_1^{{\rm{PT}}},...,x_N^{{\rm{PT}}}} \kappa {\left| {\sum\limits_{n \in {\cal N}} {\frac{{{e^{ - j{\phi _{n,s}}}}}}{{{{\left\| {{{\bf{u}}_s} - {\bf{p}}_n^{{\rm{PT}}}} \right\|}^{\frac{\chi }{2}}}}}} } \right|^2} + \sum\limits_{n \in {\cal N}} {\frac{1}{{{{\left\| {{{\bf{u}}_s} - {\bf{p}}_n^{{\rm{PT}}}} \right\|}^\chi }}}} 
			\end{aligned}
			\label{eq15_a}
		\end{equation}
		\begin{equation}
			\begin{aligned}
				{\mathrm{ s.t.}} \ \eqref{eq14_4},  \eqref{eq14_6},
			\end{aligned}
			\label{eq15_b}
		\end{equation}
	\end{subequations}
	where ${\phi _{n,r}} = 2\pi \left( {\frac{{\left\| {{{\bf{u}}_r} - {\bf{p}}_n^{{\rm{PT}}}} \right\|}}{\lambda } + \frac{{\left\| {{\bf{p}}_0^{{\rm{PT}}} - {\bf{p}}_n^{{\rm{PT}}}} \right\|}}{{{\lambda _g}}}} \right)$ represents the cumulative phase shift resulting from the $n$-th primary PA to user $r$ propagation both in free space and within the waveguide.

	Note that the objective function couples large-scale path-loss effects with wavelength-level phase adjustments, which makes the optimization landscape highly sensitive to small perturbations in primary PAs placement. These factors make a global joint optimization prohibitively complex, thereby motivating the hierarchical decomposition strategy. For problem (\rm P2), the impact of the primary PA positions arises from two mechanisms: (i) waveguide-level large-scale free-space path loss, and (ii) wavelength-level phase perturbations that shape the aggregate phase response and hence the coherent combining or cancellation of the received signals.
	
	Furthermore, the positional adjustments at the wavelength level are substantially smaller than those made at the waveguide level. Consequently, their effect on the large-scale path loss can be disregarded in the analysis. This leads to distinct optimization principles for the signal and interference terms. Maximizing the desired signal gain ${\psi _p^{{\rm{PT}}}}$, necessitates the joint optimization of both the signal amplitude and coherent phase alignment to achieve constructive superposition at the PU. In contrast, the suppression of the interference term, ${\psi _s^{{\rm{PT}}}}$, is primarily achieved through precise phase configuration to induce destructive superposition, rather than direct amplitude minimization. Note that even if the individual interference amplitudes are large, they can be effectively canceled at the SU by configuring their phases to be antagonistic, such that their complex vectors sum to near zero. Therefore, the optimization prioritizes precise phase configuration to achieve this cancellation, rendering the direct minimization of the path-loss amplitudes unnecessary.
	
	This structural insight motivates a tractable, two-step optimization framework that addresses the waveguide-level and wavelength-level designs sequentially. First, a waveguide-level placement of the primary PAs to minimize the aggregate free-space path loss and maximize the effective channel gain at the PU. Second, a wavelength-level adjustment of the primary PA positions to enforce constructive phase alignment at the PU while inducing destructive combining at the SU to suppress interference. The wavelength-scale repositioning of the PAs are several orders of magnitude smaller than the coarse waveguide-level placements. Consequently, the PT to PU link distance and the associated large-scale path-loss term are considered invariant during the phase-domain optimization.
	
	\subsubsection{Coarse-Scale (Waveguide-Level) Optimization} \label{PTPA_OP2}
	Based on the aforementioned theoretical framework, problem (\rm P2) can be addressed by first solving the following path loss minimization problem
	\begin{subequations}
		\begin{equation}
			\left( {{\rm{P2}}.1} \right) \mathop {\min }\limits_{x_1^{{\rm{PT}}},...,x_N^{{\rm{PT}}}} \sum\limits_{n \in \mathcal N} {{{\left[ {{{\left( {x_n^{{\rm{PT}}} - {x_p}} \right)}^2} + C_p^{{\rm{PT}}}} \right]}^{\frac{\chi }{4}}}} 
			\label{eq:obj}
		\end{equation}
		\begin{equation}
			{\mathrm{ s.t.}} \ {\eqref{eq14_4}},  \eqref{eq14_6}.
		\end{equation}
	\end{subequations}
	
	\textit{Lemma}~3:
	For the path-loss minimization problem (P2.3), the optimal placement of primary PAs necessarily satisfies the inter-element spacing constraint with equality, which is denoted as
			\begin{equation}
				x_n^{\rm PT*} - x_{n-1}^{\rm PT*} = \Delta_{\min}, \quad n \in \mathcal{N}.
				\label{eq:spacing_active}
			\end{equation}
	
	\textit{Proof:} The proof is analogous to that in Appendix~A of \cite{PA-carefully2} and is therefore omitted for conciseness.
	
	Once the optimal location of the first primary PA $x_1^{\rm PT}$ is determined, the positions of the subsequent primary PAs can be sequentially determined and are given by
	\begin{equation}
		x_n^{\rm PT} = x_1^{\rm PT} + (n-1)\Delta_{\min},\quad n \in \mathcal N.
		\label{eq:equidistant_positions}
	\end{equation}
	
	Accordingly, the original multi-variable objective function can be reformulated as a univariate function with respect to $x_1^{\rm PT}$
	\begin{equation}
		g(x_1^{\rm PT}) = \sum\limits_{n = 1}^N {{{\left[ {{{(x_1^{{\rm{PT}}} + (n - 1){\Delta _{\min }} - {x_p})}^2} + C_p^{{\rm{PT}}}} \right]}^{\frac{\chi }{4}}}},
		\label{eq:g_x1}
	\end{equation}
	where $C_r^{{\rm{PT}}} = {\left( {{y^{{\rm{PT}}}} - {y_r}} \right)^2} + {\left( {l - {z_r}} \right)^2}$ is the aggregated squared distance in the $y$ and $z$ dimensions.	The function $g(x_1^{\rm PT})$ is unimodal in $x_1^{\rm PT}$ because each term is convex and their sum preserves unimodality. For sufficiently large $C_p^{\rm PT}$, the function $g(x_1^{\rm PT})$ is strictly convex, ensuring that the stationary point in~\eqref{eq:center_solution} is the unique global minimizer. To determine this minimizer, we set the first derivative of $g(x_1^{\rm PT})$ with respect to $x_1^{\rm PT}$ equal to zero, which yields
	\begin{equation}
		x_1^{\rm PT*} =
		\begin{cases}
			x_p - \dfrac{N-2}{2}\Delta_{\min}, & \text{if $N$ is even}, \\[10pt]
			x_p - \dfrac{N-1}{2}\Delta_{\min}, & \text{if $N$ is odd}.
		\end{cases}
		\label{eq:center_solution}
	\end{equation}
	
	As rigorously established in~\cite{PA-carefully2}, the objective function is strictly decreasing for $x_1^{\rm PT} \in [x^{\rm PT}_{\min}, x_1^{\rm PT*}]$ and strictly increasing for $x_1^{\rm PT} \in [x_1^{\rm PT*}, x^{\rm PT}_{\max}]$. This monotonicity ensures that the stationary point in~\eqref{eq:center_solution} is the unique global minimizer. Therefore, $x_1^{\rm PT*}$ achieves the minimum aggregate path loss for the primary PAs, regardless of whether $N$ is odd or even. Accordingly, the $n$-th primary PA position is given by
	\begin{equation}
		x_n^{\rm PT*} =
		\begin{cases}
			x_p - \frac{N-2}{2}\Delta_{\min} + (n-1)\Delta_{\min}, & \text{if $N$ is even}, \\[10pt]
			x_p - \frac{N-1}{2}\Delta_{\min} + (n-1)\Delta_{\min}, & \text{if $N$ is odd}.
		\end{cases}
		\label{eq:optimal_positions1}
	\end{equation}
	
	\begin{figure}[tpb]
	\centering
	\includegraphics[width=0.95\linewidth]{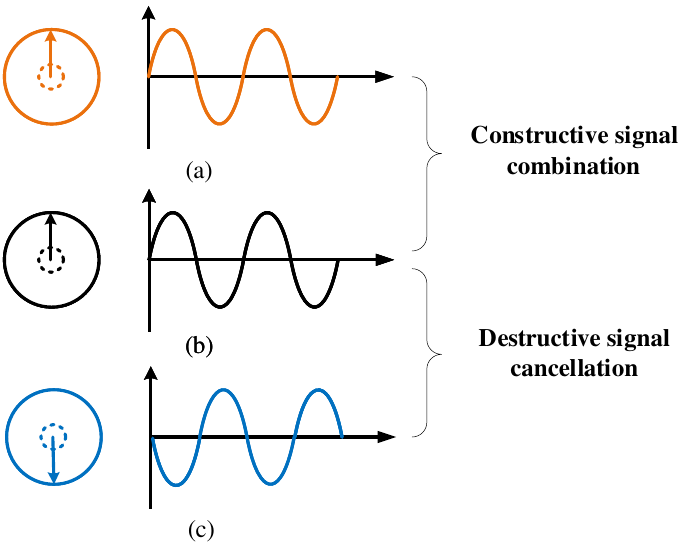}
	\caption{Illustration of the received signal at the user under phase alignment and phase cancellation schemes: (a) $(t{+}1)$-th PA with phase alignment; (b) $t$-th PA; (c) $(t{+}1)$-th PA with phase cancellation.}
	\label{Fig2}
\end{figure}
	\subsubsection{Fine-Scale (Wavelength-Level) Optimization} \label{PTPA_OP3} With the determined macroscopic locations, the next step is to perform the fine-scale optimization. Specifically, the phase shift from $n$-th primary PA to the PU and SU can be expressed as
	\begin{equation}
		{\phi _{n,p}} = \frac{{2\pi }}{\lambda }\sqrt {{{\left( {x_n^{{\rm{PT}}} - {x_p}} \right)}^2} + C_p^{{\rm{PT}}}}  + \frac{{2\pi }}{{{\lambda _g}}}\left( {x_n^{{\rm{PT}}} - x_0^{{\rm{PT}}}} \right),
		\label{eq18}
	\end{equation}
	\begin{equation}
		{\phi _{n,s}} = \frac{{2\pi }}{\lambda }\sqrt {{{\left( {x_n^{{\rm{PT}}} - {x_s}} \right)}^2} + C_s^{{\rm{PT}}}}  + \frac{{2\pi }}{{{\lambda _g}}}\left( {x_n^{{\rm{PT}}} - x_0^{{\rm{PT}}}} \right).
	\end{equation}
	
	As depicted in Fig.~\ref{Fig2}(a) and (b), maximizing the received signal strength at the PU requires a precise adjustment of the primary PA positions to ensure constructive phase alignment at the PU. Accordingly, the constructive signal combination strategy can be mathematically formulated as
	\begin{equation}
		\begin{aligned}
			{\phi _{n,p}} - {\phi _{n - 1,p}} = 2{k_1}\pi , n \in {\mathcal N},
		\end{aligned}
		\label{eq16_d}
	\end{equation}
	where $k_1 >0$ is a positive integer constant that ensures the phase shifts between adjacent antennas differ by integer multiples of $2 \pi$.
	
	As illustrated in Figs.~\ref{Fig2}(a) and (c), effective signal cancellation at the SU from the signals emitted by the primary PAs requires consideration of two distinct cases, depending on the number of primary PAs. The objective is to ensure that the signals from the primary PAs are mutually canceled at the SU, thereby minimizing interference and optimizing system performance.
	
	$\bullet$ \textit{Even Number of primary PAs}: When the number of primary PAs is even, perfect destructive interference at the SU can be achieved by configuring the phase difference between adjacent primary PAs to be $\pi$, ensuring that the signals combine in anti-phase and cancel each other. The corresponding destructive interference cancellation scheme is given by
	\begin{equation}
		{\phi _{n,s}} - {\phi _{n - 1,s}} = {\rm{  }}2{k_2}\pi  + \pi ,n \in {\cal N},
		\label{EQ29}
	\end{equation}
	where $k_2 \in {\mathbb Z}^+$ is a positive integer. The constraint (\ref{EQ29}) ensures that for every pair of adjacent primary PAs, the phase shift between them induces complete cancellation at $u_s$. 
	
		\begin{figure}[htpb]
		\centering
		\includegraphics[width=3.5in]{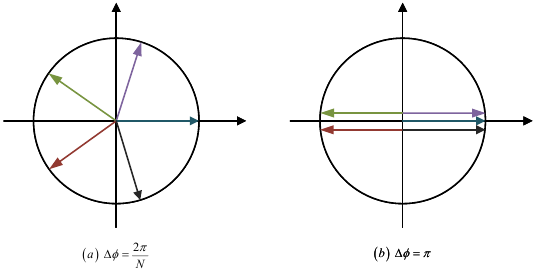}
		\caption{Illustration of complex vector addition for $N=5$ unit phasors with (a) Uniform phase spacing: $\Delta\phi = 2\pi/N$ (b) Alternating phase: $\Delta\phi = \pi$. }
		\label{Fig_odd}
	\end{figure} 
	
	$\bullet$ \textit{Odd Number of primary PAs}: In scenarios where the number of primary PAs is odd, the phase alignment strategy (\ref{EQ29}) adopted for even-element arrays is inadequate for achieving complete destructive superposition at the SU. As illustrated in Fig.~\ref{Fig_odd}, this limitation stems from the absence of a symmetric counterpart for the remaining antenna element that cannot be paired, which precludes pairwise phase cancellation and leads to a residual interference term. This residual interference effect is particularly pronounced when the interference between the PN and SN is strong or when the number of primary PAs is relatively small, as the lack of complete pairwise cancellation becomes more significant under these conditions. 
	
	\textit{Lemma~4}: The optimal destructive interference cancellation scheme for odd $N$ is refined as
	\begin{equation}
		\begin{aligned}
			{\phi _{n,s}} - {\phi _{n - 1,s}} = \left( {{k_2} + \frac{1}{N}} \right)2\pi ,n \in {\mathcal N}.
		\end{aligned}
		\label{eq16_e}
	\end{equation}
	
	\textit{Proof}: A comprehensive derivation of this condition is provided in Appendix~B.	
	
	In order to preserve the accuracy of the phase approximation while maintaining the optimized path loss, only minor adjustments can be applied to the positions of adjacent primary PAs. As a result, ${\phi _{n,p}}$ can be approximated using the first-order Taylor expansion around the fixed position $x_{n - 1}^{{\rm{PT}}}$, which can be expressed as 
		\begin{equation}
			\begin{aligned}
				\begin{aligned}
					{\phi _{n,p}} \approx & \ \frac{{2\pi }}{{{\lambda _g}}}\left( {x_n^{{\rm{PT}}} - x_0^{{\rm{PT}}}} \right) + \frac{{2\pi }}{\lambda }\sqrt {{{\left( {x_{n - 1}^{{\rm{PT}}} - {x_p}} \right)}^2} + C_p^{{\rm{PT}}}}  \\
					+  &  \frac{{2\pi }}{\lambda }\frac{{x_{n - 1}^{{\rm{PT}}} - {x_p}}}{{\sqrt {{{\left( {x_{n - 1}^{{\rm{PT}}} - {x_p}} \right)}^2} + C_p^{{\rm{PT}}}} }}\left( {x_n^{{\rm{PT}}} - x_{n - 1}^{{\rm{PT}}}} \right).
				\end{aligned}
			\end{aligned}
			\label{eq20}
		\end{equation}
	
	To achieve simultaneous constructive signal combination at the PU and destructive interference cancellation at the SU, the incremental antenna displacement of adjacent primary PAs $\Delta x = x_n^{\mathrm{PT}} - x_{n-1}^{\mathrm{PT}}$ is determined by jointly solving the phase alignment constraints at both users. Specifically, by substituting the first-order Taylor approximation (\ref{eq20}) into constraints~\eqref{eq16_d} and~\eqref{EQ29}-\eqref{eq16_e}, we obtain
	\begin{equation}
		\left[
		\frac{1}{\lambda}
		\frac{x_{n-1}^{\mathrm{PT}} - x_p}{\sqrt{(x_{n-1}^{\mathrm{PT}} - x_p)^2 + C_p^{\mathrm{PT}}}}
		+ \frac{1}{\lambda_g}
		\right]
		\Delta x = k_1,
		\label{eq:PU_constraint}
	\end{equation}
	and	
	\begin{align}
		& \left[
		\frac{1}{\lambda}
		\frac{x_{n-1}^{\mathrm{PT}} - x_s}{\sqrt{(x_{n-1}^{\mathrm{PT}} - x_s)^2 + C_s^{\mathrm{PT}}}}
		+ \frac{1}{\lambda_g}
		\right]
		\Delta x 
		= b(k_2),
		\label{eq:SU_constraint}
	\end{align}
	where 
	\begin{equation}
		b(k_2) =
		\begin{cases}
			k_2 + 0.5, & \text{if } N \text{ is even,} \\[4pt]
			k_2 + \dfrac{1}{N}, & \text{if } N \text{ is odd,}
		\end{cases}
		\label{eq:b_definition}
	\end{equation}
	encapsulates the phase increment requirement at the SU for destructive combining of the primary PA signals.
	
	Let $\mathcal{K}_1$ and $\mathcal{K}_2$ denote the feasible integer search ranges for $k_1$ and $k_2$, respectively (e.g., ${{\cal K}_1} = {\rm{ }}\left\{ {1, \ldots ,{K_{\max }}} \right\}$). For each $k_1\in\mathcal{K}_1$ and $k_2\in\mathcal{K}_2$, we compute the candidate step sizes as
		\begin{equation}
			\Delta x_1(k_1) =
			\frac{k_1}{
				\frac{1}{\lambda}
				\frac{x_{n-1}^{\mathrm{PT}} - x_p}{\sqrt{(x_{n-1}^{\mathrm{PT}} - x_p)^2 + C_p^{\mathrm{PT}}}}
				+ \frac{1}{\lambda_g}
			},
			\label{eq:delta_x1}
		\end{equation}
		\begin{equation}
			\Delta x_2(k_2) =
			\frac{b(k_2)}{
				\frac{1}{\lambda}
				\frac{x_{n-1}^{\mathrm{PT}} - x_s}{\sqrt{(x_{n-1}^{\mathrm{PT}} - x_s)^2 + C_s^{\mathrm{PT}}}}
				+ \frac{1}{\lambda_g}
			}.
			\label{eq:delta_x2}
		\end{equation}
	
	The objective is to identify the pair $(k_1^*,k_2^*)$ such that the absolute difference between $\Delta x_1(k_1)$ and $\Delta x_2(k_2)$ is minimized, which can be expressed as
	\begin{equation}
		(k_1^*,k_2^*) = \arg\min_{k_1\in\mathcal{K}_1,k_2\in\mathcal{K}_2} \left| \Delta x_1(k_1) - \Delta x_2(k_2) \right|.
		\label{eq38}
	\end{equation}
	
	The optimal incremental displacement is then selected as
	\begin{equation}
		\Delta x^* = \Delta x_1(k_1^*),
		\label{EQ39}
	\end{equation}
	thereby prioritizing strict phase coherence at the PU, which is typically of paramount importance in CR networks.
	
	Subsequently, the position of the $n+1$-th primary PA is recursively updated according to
	\begin{equation}
		x_{n+1}^{\mathrm{PT}} = x_{n}^{\mathrm{PT}} + \Delta x^*.
		\label{popopo}
	\end{equation}
	
	Alternatively, for a given ${x_{n}^{{\rm{PT}}}}$, we can similarly refine ${x_{n - 1}^{{\rm{PT}}}}$.
	
	This joint search procedure guarantees that the spatial configuration of the primary PA achieves a near-optimal trade-off between constructive signal aggregation at the PU and interference suppression at the SU, under the practical constraints imposed by discrete phase increments and physical deployment limitations. Nevertheless, it should be noted that the joint phase increment search will inevitably introduce a residual phase error, which is fundamentally attributed to the finite resolution of the discrete search space and the theoretical approximations adopted in the algorithmic design. 	
	
	\textit{Remark~1:}
	The residual phase error arising from the joint phase increment search is fundamentally governed by three factors. First, the integer search spaces for $(k_1, k_2)$ are necessarily finite, ensuring that the incremental displacement $\Delta x$ remains sufficiently small such that the large-scale path loss between the primary PA and the PU can be approximated as uniform across the antennas, i.e., $\left| x_n^{\mathrm{PT}} - x_{n-1}^{\mathrm{PT}} \right| \ll \left| x_n^{\mathrm{PT}} - x_p \right|$. Excessively enlarging the search range would violate this approximation, thus compromising the analytical decoupling between spatial and phase optimization. Second, the validity of the first-order Taylor expansion employed in (\ref{eq20}) presumes that $\Delta x$ is infinitesimal relative to the propagation distance. Otherwise, higher-order effects become non-negligible. Third, the algorithm prioritizes exact phase alignment at the PU by setting $\Delta x^* = \Delta x_1(k_1^*)$, which inherently leads to imperfect destructive interference at the SU due to the structure of the coupled constraints. The minimum residual step mismatch is
	\begin{equation}
		\Delta \varepsilon = \min_{k_1, k_2} \left| \Delta x_1(k_1^*) - \Delta x_2(k_2^*) \right|.
	\end{equation}
	
	For $b(k_2) = k_2 +0.5$, the induced phase mismatch at the SU is generally negligible. In contrast, for $b(k_2) = k_2+ \dfrac{1}{N}$ (especially for large $N$), the same $\Delta \varepsilon$ is much larger than the phase increment, leading to a non-negligible deviation from ideal interference cancellation.
	
	To address this limitation in odd-numbered primary PAs, a refined phase assignment strategy is proposed. Specifically, for the three antennas located around the array center (i.e., the $(\frac{N + 1}{2} - 1)$-th, $(\frac{N + 1}{2})$-th, and $(\frac{N + 1}{2} + 1)$-th primary PAs), the phase differences are assigned as
	\begin{equation}
		{\phi _{n,s}} - {\phi _{n - 1,s}} = 2{k_2}\pi  + \frac{{2\pi }}{3},n \in \left\{ {\frac{{N + 1}}{2},\frac{{N + 1}}{2} + 1} \right\},
		\label{PTPA_SC}
	\end{equation}
	while for the remaining even-numbered primary PAs, the phase difference is maintained at \(2\pi\) to ensure mutual cancellation. This strategy simultaneously resolves two critical limitations: (i) it suppresses the residual interference inherent in odd-numbered primary PAs resulting from incomplete pairwise phase cancellation-an effect that is particularly evident for small array sizes; and (ii) it alleviates the sensitivity to phase errors under the interference cancellation condition in~(\ref{eq16_e}), whose impact becomes increasingly significant as the number of primary PAs grows.
	
		\subsection{Secondary Pinching Beamforming Optimization Design}
		\label{STPA_OP}
		\subsubsection{Problem Reformulation}
		When the variables $\left( {{p_{\rm ST}},{{\bf{x}}^{{\rm{PT}}}}} \right)$ are given, the problem (\rm P1) can be equivalently reformulated as
		
		\begin{subequations}
			\begin{equation}
				\begin{aligned}
					\left( {{\rm{P3}}} \right)\;\mathop {\max }\limits_{{{\bf{x}}^{{\rm{ST}}}}} \;\psi _s^{{\rm{ST}}}\left( {{{\bf{x}}^{{\rm{ST}}}}} \right){\rm{ and }}\;\mathop {\min }\limits_{{{\bf{x}}^{{\rm{ST}}}}} \psi _p^{{\rm{ST}}}\left( {{{\bf{x}}^{{\rm{ST}}}}} \right)
				\end{aligned}
				\label{eq15_aa}
			\end{equation}
			\begin{equation}
				\begin{aligned}
					{\mathrm{ s.t.}} \ \eqref{eq14_3}, \eqref{eq14_4},  \eqref{eq14_6}.
				\end{aligned}
				\label{eq15_bb}
			\end{equation}
		\end{subequations}
		
		It is readily observed that the formulation of problem (\rm P3) is similar to problem (\rm P2), except for the inclusion of the ITC~(\ref{eq14_3}), which restricts the aggregate interference imposed upon the PU within a prescribed threshold.
		
		\textit{Remark~2:}
		The phase-aligned interference cancellation strategy introduced in Section~\ref{PTPA_OP2} substantially mitigates the interference at the PU, such that the ITC requirement in problem (P3) is typically non-binding. This observation can be attributed to: i) the phase coordination scheme reinforces the desired signal at the SU while enforcing destructive superposition at the PU, thereby driving the interference level well below the admissible threshold, and ii) any residual interference after phase suppression is further weakened by large-scale path loss, since the secondary PAs are inherently deployed closer to the SU than to the PU. Moreover, the optimization framework explicitly retains the ITC constraint in (P4) to ensure reliable PU protection under all channel conditions. In particular, during the ST power allocation stage, the optimization inherently regulates the transmit power within the admissible range, thereby guaranteeing compliance with the ITC while preserving the robustness of the overall design.
		
		Therefore, the optimization of secondary PA locations can follow an analogous approach to the primary PA scenario, which is given as follows.
		
		\subsubsection{Coarse-Scale (Waveguide-Level) Optimization for Secondary PA Locations}	
		In direct analogy to the optimization of the primary PA, the initial stage of secondary PA deployment seeks to minimize the cumulative path loss from the secondary PAs to the SU, which can be formally expressed as
		\begin{subequations}
			\begin{equation}
				(\mathrm{P3.1})\quad \min_{x_1^{\rm ST}, \ldots, x_M^{\rm ST}} \sum\limits_{m = 1}^M {\left[ {{{(x_m^{{\rm{ST}}} - {x_s})}^2} + C_s^{{\rm{ST}}}} \right]^{\frac{\chi }{4}}}
				\label{eq:stpa_obj}
			\end{equation}
			\begin{equation}
				{\mathrm{ s.t.}} \ \eqref{eq14_4}, \eqref{eq14_6},
				\label{eq:stpa_constraints}
			\end{equation}
		\end{subequations}
		where $C_s^{\rm ST}$ denotes the aggregate squared distances in the $y$ and $z$ dimensions.
		
		Similarly, the closed-form expression for the initial secondary PA position is
		\begin{equation}
			x_1^{\rm ST*} =
			\begin{cases}
				x_s - \dfrac{M-2}{2}\Delta_{\min}, & \text{if } M \text{ is even} \\[10pt]
				x_s - \dfrac{M-1}{2}\Delta_{\min}, & \text{if } M \text{ is odd,}
			\end{cases}
			\label{eq:stpa_center}
		\end{equation}
		where the location of the $m$-th secondary PA given by
		\begin{equation}
			x_m^{\rm ST*} = x_1^{\rm ST*} + (m-1)\Delta_{\min}, \quad m \in {\mathcal M}.
			\label{eq:stpa_optimal_positions}
		\end{equation}
		
		\subsubsection{Fine-Scale (Wavelength-Level) Optimization for Secondary PA Locations}
		Based on the macroscopic deployment determined by (\ref{eq:stpa_center}) and (\ref{eq:stpa_optimal_positions}), a subsequent microscopic refinement is performed to ensure constructive phase alignment at the SU and destructive interference at the PU. Following the same procedure from \eqref{eq18}-\eqref{eq:SU_constraint}, the candidate step sizes are denoted as
		\begin{equation}
			\Delta x_3(k_3) =
			\frac{k_3}{
				\frac{1}{\lambda}
				\frac{x_{m-1}^{\rm ST} - x_s}{\sqrt{(x_{m-1}^{\rm ST} - x_s)^2 + C_s^{\rm ST}}}
				+ \frac{1}{\lambda_g}
			},
			\label{eq:stpa_dx1}
		\end{equation}
		\begin{equation}
			\Delta x_4(k_4) =
			\frac{b(k_4,m)}{
				\frac{1}{\lambda}
				\frac{x_{m-1}^{\rm ST} - x_p}{\sqrt{(x_{m-1}^{\rm ST} - x_p)^2 + C_p^{\rm ST}}}
				+ \frac{1}{\lambda_g}
			},
			\label{eq:stpa_dx2}
		\end{equation}
		where $k_3, k_4 \in \mathbb{Z}_+$ denote integer-valued phase indices, while
		\begin{equation}
			b(k_4, m) =
			\begin{cases}
				k_4 + \dfrac{1}{3}, & \text{if } M \text{ is odd and } m \in \mathcal{M}_c \\[12pt]
				k_4 + 0.5, & \text{otherwise,}
			\end{cases}
			\label{eq:stpa_b_def}
		\end{equation}
		characterizes the phase increment requirement for destructive combining at the PU, with $\mathcal{M}_c = \left\{ \frac{M+1}{2},\; \frac{M+3}{2} \right\}$.
		
		The optimal displacement is then determined by identifying the pair $(k_3^*, k_4^*)$ that minimizes the difference 
		\begin{equation}
			(k_3^*,k_4^*) = \arg\min_{k_3 \in \mathcal{K}_3, k_4 \in \mathcal{K}_4} \left| \Delta x_3(k_3) - \Delta x_4(k_4) \right|.
			\label{STPAP}
		\end{equation}
		\begin{equation}
			x_m^{\rm ST} = x_{m-1}^{\rm ST} + \Delta x^*,
			\label{ssss}
		\end{equation}
		and selecting $\Delta x^* = \Delta x_3(k_3^*)$ to prioritize strict phase coherence at the SU.
		
		Alternatively, for a prescribed $x_m^{\rm ST}$, the previous antenna position may be updated as $x_{m-1}^{\rm ST} = x_m^{\rm ST} - \Delta x^*$.
		
		\subsection{ST Power Control} 
		Due to the effective interference cancellation at the PU achieved by the proposed pinching beamforming optimization design, the ASE of the PU remains essentially invariant with respect to the secondary PA transmit power. Consequently, in the subsequent power allocation step, it is reasonable to focus solely on maximizing the SU's ASE, thereby simplifying the optimization process and enhancing computational efficiency. For given locations of the primary and secondary PAs, the problem (P1) can be approximately transformed into
		\begin{subequations}
			\begin{equation}
				\begin{aligned}
					\left( {{\rm{P4}}} \right) \ \mathop {\max }\limits_{{p_{\rm ST}}} \ {\log _2}\left( {1 + \frac{{N{p_{\rm ST}}\psi _s^{{\rm{ST}}}}}{{M{P_{{\rm{PT}}}}\psi _s^{{\rm{PT}}} + NM{\sigma ^2}}}} \right)
				\end{aligned}
				\label{eq15_1}
			\end{equation}
			\begin{equation}
				\begin{aligned}
					{\mathrm{ s.t.}} \ 0 < {p_{\rm ST}} \le \min \left( {{P_{{\rm{ST}}}},p_{\rm ST}^{{\rm{TH}}}} \right),
				\end{aligned}
				\label{eq15_2}
			\end{equation}
		\end{subequations}
		where ${\psi _s^{{\rm{ST}}}}$ and ${\psi _s^{{\rm{PT}}}}$ are constants for given $\left( {{{\bf{x}}^{{\rm{PT}}}},{{\bf{x}}^{{\rm{ST}}}}} \right)$, and $p_{\rm ST}^{{\rm{TH}}} = \frac{{M{P^{{\rm{TH}}}}}}{{\psi _p^{{\rm{ST}}}}}$ denotes the maximum transmit power that satisfies the ITC. It is evident that the objective function is monotonically increasing with respect to $p_{\rm ST}$. Consequently, the maximum SE is achieved with the optimized ST transmission power
		\begin{equation}
			p_{\rm ST}^* = \min \left( {{P_{{\rm{ST}}}},p_{\rm ST}^{{\rm{TH}}}} \right),
		\end{equation}
        which provides a closed-form solution for the ST power control problem under the considered interference constraint.	
		
			\begin{algorithm}[t] 
			\caption{\textbf{Proposed}: Three-Stage Joint Optimization Algorithm}
			\label{alg:pcm-jpo-once}
			\begin{algorithmic}[1]
				\STATE \textbf{Initialize:} Feasible solutions ${\bf x}^{\mathrm{PT}}$, $\bf{x}^{\mathrm{ST}}$, $p_{\rm ST}$, and integer bound $K_{\max}$.
				
				\STATE \textbf{Stage 1: Primary pinching beamforming optimization}
				\STATE \hspace{\algorithmicindent} \textbf{Stage 1.1 (Coarse):} Minimize path loss toward the
				\hspace*{0.19in}PU and place the primary PA locations using (\ref{eq:optimal_positions1}) with  
				\hspace*{0.19in}equal inter-element spacing.
				\STATE \hspace{\algorithmicindent} \textbf{Stage 1.2 (Refine):} Enforce constructive alignment 
				\hspace*{0.19in}at PU by (\ref{eq16_d}) and destructive combining at SU using 
				\hspace*{0.19in}(\ref{EQ29}) for even $N$, or apply the center-triplet rule per 
				\hspace*{0.19in}(\ref{PTPA_SC}) for odd $N$. Convert phase constraints to displac-\hspace*{0.18in}ement candidates $\Delta x_1(k_1)$ and $\Delta x_2(k_2)$, then solve 
				\hspace*{0.19in}(\ref{eq38}) and set $\Delta x^\star=\Delta x_1(k_1^\star)$ to maintain PU coher-\hspace*{0.18in}ence. Update primary PA positions recursively using 
				\hspace*{0.13in} (\ref{popopo}) to obtain $\bf{x}^{{\mathrm{PT}}\star}$.
				
				\STATE \textbf{Stage 2: Secondary pinching beamforming optimization}
				\STATE \hspace{\algorithmicindent} \textbf{Stage 2.1 (Coarse):} Minimize path loss toward SU \hspace*{0.17in} with equal spacing and closed-form position from (\ref{eq:stpa_optimal_positions}), 
				\hspace*{0.20in}then fill elements sequentially.
				\STATE \hspace{\algorithmicindent} \textbf{Stage 2.2 (Refine):} Enforce constructive alignment \hspace*{0.20in}at SU and destructive combining at PU. Calculate 
				\hspace*{0.20in}$\Delta x_3(k_3)$ and $\Delta x_4(k_4)$ based on (\ref{STPAP}). Set $\Delta x^\star  = \hspace*{0.16in} \Delta x_3(k_3^\star)$ according to (\ref{ssss}) to obtain $\bf{x}^{{\mathrm{ST}}\star}$.
				
				\STATE \textbf{Stage 3: ST power control}
				\STATE \hspace{\algorithmicindent} Compute $p_{\rm ST}^{\star}=\min\!\big(P_{\mathrm{ST}},\, p_{\rm ST}^{\mathrm{TH}}\big)$ solving (P4), ensur
				\hspace*{0.19in}-ing the interference temperature constraint is satisfied.
				
				\STATE \textbf{Output:} The optimized $\bf{x}^{{\mathrm{PT}}\star}$, $\bf{x}^{{\mathrm{ST}}\star}$, $p_{\rm ST}^{\star}$.
			\end{algorithmic}
		\end{algorithm}
		
		\subsection{Computational Complexity}	
		The complete execution procedure of the proposed algorithm is outlined in \textbf{Algorithm 1}. Denoting $K_{\max}$ as  the integer search bound in the refinement steps, the computational complexity for optimizing the primary PA placement is $O(NK_{\max}^{2})$. Similarly, solving problem (P3) to determine the optimized secondary PA placement entails a complexity of $O(MK_{\max}^{2})$, while solving problem (P4) for the ST transmit power requires $O(1)$ operations. Consequently, the overall complexity per run is $O(\left( {N + M} \right)K_{\max}^{2})$.

		\section{Simulation Results} \label{sec:simulation}
		In this section, numerical simulations are performed to demonstrate the effectiveness and superiority of the PA-enabled CR network and the proposed optimization algorithm.			

		\subsection{Simulation Setup and Comparative Algorithms}	
		Unless otherwise specified, the main simulation parameters are summarized as follows. The carrier frequency is $f = 28~\mathrm{GHz}$, and the effective relative permittivity of the waveguide is chosen as $\eta_{\mathrm{eff}} = 1.4$. The minimum antenna spacing is set to $\Delta_{\mathrm{min}} = \lambda/2$. The Ricean factor and the path loss exponent are configured as $\kappa = 4$ and $\chi =2.2$, respectively. The numbers of primary and secondary PAs are set as $N=M=5$, respectively. The waveguide dimensions are $L_x = 15~\mathrm{m}$ in length and $l = 3~\mathrm{m}$ in height, respectively. The separation between the PT and ST along the $y$-axis is $d = 12~\mathrm{m}$. The PT and ST waveguides are symmetrically placed about the origin along the $y$-axis, and thus $y^{\mathrm{ST}} = -y^{\mathrm{PT}} = \frac{d}{2}$, and their feed points are aligned at $x_0^{\mathrm{PT}} = x_0^{\mathrm{ST}} = 0~\mathrm{m}$. The PU and SU are randomly and independently distributed within the simulation region, with the PU coordinates drawn from $x_p \sim \mathcal{U}(0, L_x)$ and $y_p \sim \mathcal{U}(y^{\mathrm{PT}} - {L_y}/{2},, y^{\mathrm{PT}} + {L_y}/{2})$, while the SU coordinates from $x_s \sim \mathcal{U}(0, L_x)$ and $y_s \sim \mathcal{U}(y^{\mathrm{ST}} - {L_y}/{2},, y^{\mathrm{ST}} + {L_y}/{2})$, where $L_y$ denotes the width of the user distribution region. Moreover, the $z$-coordinates of both users are fixed at $0$~m and $P_{\mathrm{PT}} =P_{\mathrm{ST}} = 0~\mathrm{dBm}$, respectively, while the noise power at both users is $\sigma^2 = -90~\mathrm{dBm}$.
    	
    	To evaluate the effectiveness of the proposed PA-enabled CR network and the corresponding joint optimization algorithm, the following benchmark schemes are considered. For clarity in simulation results, we refer to our proposed algorithm as {\textbf{Proposed}}. 
\begin{itemize}				
	\item \textbf{{Ideal Case}}: This benchmark performs solely constructive beamforming optimization aimed at signal enhancement at the intended receiver, while artificially suppressing the interference power at the unintended receiver to zero. The ideal sum SE can be expressed as
	\begin{equation}
		\begin{aligned}
			{R^{{\rm{Ideal}}}} &= {\log _2}\left[ {1 + \frac{{{P_{{\rm{PT}}}}\max \left( {\psi _p^{{\rm{PT}}}} \right)}}{{N{\sigma ^2}}}} \right]\;\\
			& + {\log _2}\left[ {1 + \frac{{{p_{{\rm{ST}}}^*}\max \left( {\psi _s^{{\rm{ST}}}} \right)}}{{M{\sigma ^2}}}} \right].
		\end{aligned} \nonumber
	\end{equation}
	
	Therefore, this benchmark serves as a theoretical benchmark that provides an upper bound on performance under the assumption of perfect interference suppression, which may not be achievable in practice.
	
	\item \textbf{{$\theta_{\mathrm{fix}}$-Fixed-Offset Interference Cancellation ($\theta_{\mathrm{fix}}$-FOC-IC)}} algorithm: These comparative algorithms enforce a constant inter-element phase shift between adjacent PAs for interference cancellation at the unintended receiver. The phase increment is denoted as
	\begin{equation}
		\phi_{t,r}-\phi_{t-1,r} \;=\; 2k\pi \;+\;\theta_{\mathrm{fix}},
		\quad {\theta _{{\rm{fix}}}} \in \left\{ {\mkern 1mu} \frac{{2\pi }}{{\left| {{{\cal N}_i}} \right|}}\;\pi \right\} {\mkern 1mu}. \nonumber
	\end{equation}
	
	As illustrated in Fig.~\ref{Fig_odd} (a), the \textbf{$\frac{2\pi}{| \mathcal{N}_i |}$-FOC-IC} algorithm operates by applying a uniform inter-element phase shift of $2\pi / |\mathcal{N}_i|$. This results in $| \mathcal{N}_i |$ complex phasors being uniformly distributed over the unit circle, thereby facilitating interference management.
	
	As shown in Fig.~\ref{Fig_odd} (b), the \textbf{\(\pi\)-FOC-IC} algorithm employs a phase shift of \(\pi\) between adjacent PAs to achieve interference cancellation at the unintended receiver, as presented in \cite{PA-PLS2}. While this configuration achieves perfect interference cancellation when the number of PAs is even, one element remains unpaired in the case of an odd number of PAs, leading to a residual interference component and thus incomplete suppression.
	
	\item \textbf{{Fixed-Position Antennas (FPA)}}: This benchmark corresponds to a conventional fixed BS enabled CR network wherein both the primary PA and secondary PAs are fixed in location. The PT employs maximum-ratio transmission (MRT) to maximize the PU SE, while the ST's beamformer $\mathbf{w}_{\mathrm{ST}}$ is determined by solving
	\begin{subequations}
		\begin{equation}
			\mathop {\max }\limits_{{{\bf{w}}_{{\rm{ST}}}}} \frac{{{{\left| {{\bf{h}}_{{\rm{ST - SU}}}^H{{\bf{w}}_{{\rm{ST}}}}} \right|}^2}}}{{{{\left| {{\bf{h}}_{{\rm{PT - SU}}}^H{{\bf{w}}_{{\rm{PT}}}}} \right|}^2} + {\sigma ^2}}}
		\end{equation}
		\begin{equation}
			{\mathrm{ s.t.}} \ {\rm{Tr}}\left( {{{\bf{w}}_{{\rm{ST}}}}{\bf{w}}_{{\rm{ST}}}^H} \right) \le P,
		\end{equation}
		\begin{equation}
			{\left| {{\bf{h}}_{{\rm{ST - PU}}}^H{{\bf{w}}_{{\rm{ST}}}}} \right|^2} \le {P ^{{\rm{TH}}}}.
		\end{equation}
	\end{subequations}
\end{itemize}	
     
     \begin{figure}[htpb]
     	\centering
     	\includegraphics[width=3.5in]{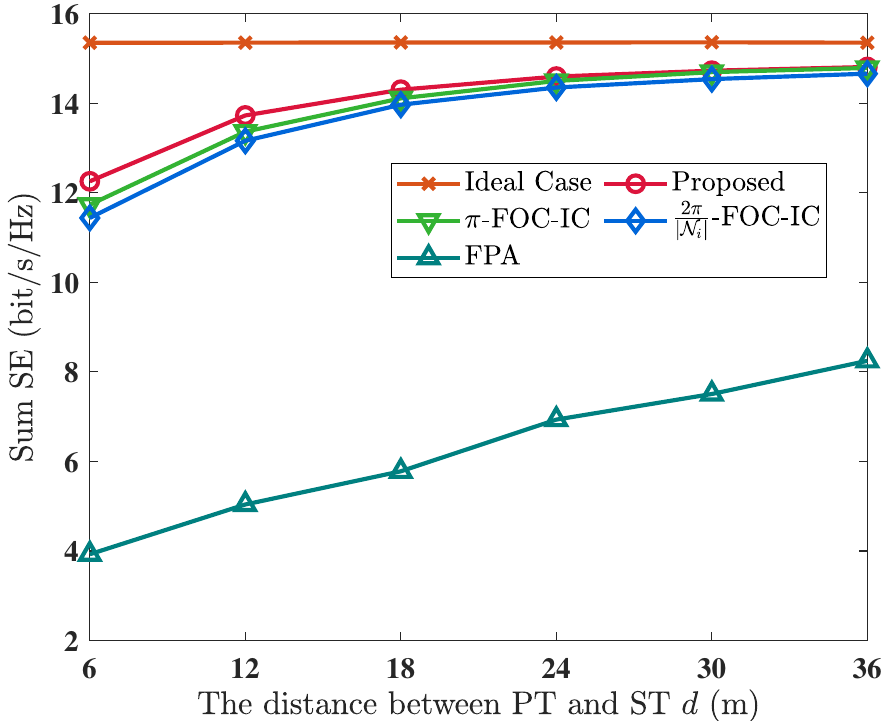}
     	\caption{Sum SE versus the distance between PT and ST.}
     	\label{fig4}
     \end{figure}
		
	\subsection{Impact of the PT-to-ST Distance}
		Figure~\ref{fig4} depicts that as the PT-to-ST distance increases, all algorithms yield improved sum SE performance which can be attributed to reduced mutual coupling among the PN and SN. Both the \textbf{Proposed} and \textbf{$\theta_{\mathrm{fix}}$-FOC-IC} algorithms outperform \textbf{FPA} algorithm, which verifies the superiority of the PA-based network architecture. Furthermore, the \textbf{Proposed} algorithm demonstrates superior performance compared to the \textbf{$\pi$-FOC-IC} and \textbf{$\frac{2\pi}{| \mathcal{N}_i |}$-FOC-IC} algorithms, particularly in severe interference scenarios where the PN is located in close proximity to the SN. This result demonstrated the effectiveness of the proposed interference cancellation scheme. Moreover, it can be observed that the performance gap between the \textbf{Proposed} algorithm and the \textbf{Ideal Case} is extremely small, particularly as the mutual distance $d$ increases. This result confirms that the \textbf{Proposed} algorithm can achieve nearly optimal sum SE performance, enabling a nearly orthogonal transmission between PN and SN.
		
		\begin{figure*}[htbp]
			\centering
			\subfigure[]{
				\includegraphics[width=0.46\linewidth]{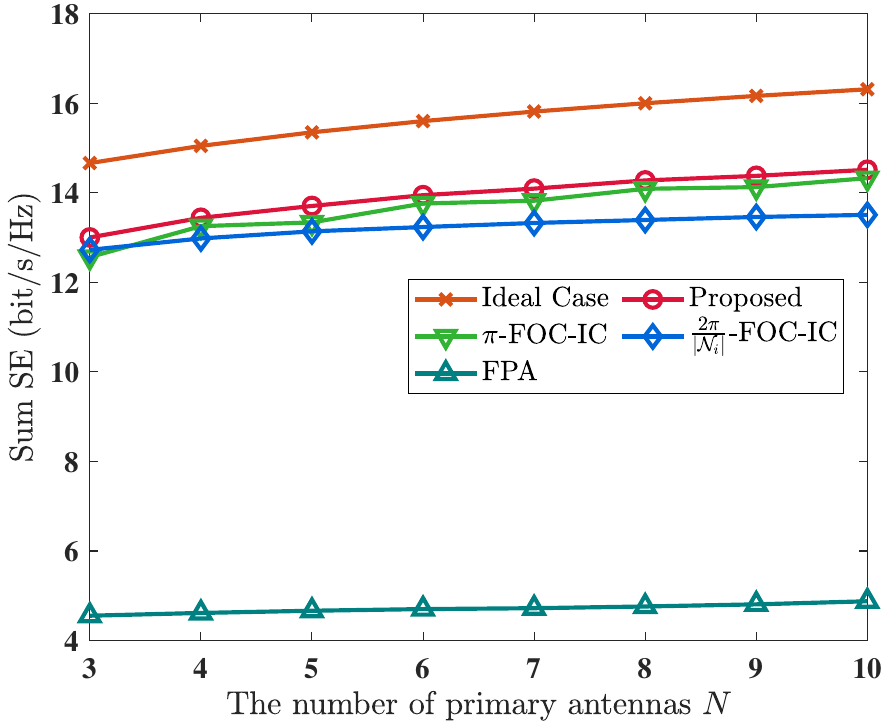}
				\label{fig5}
			}
			\hspace{0.03\linewidth}
			\subfigure[]{
				\includegraphics[width=0.46\linewidth]{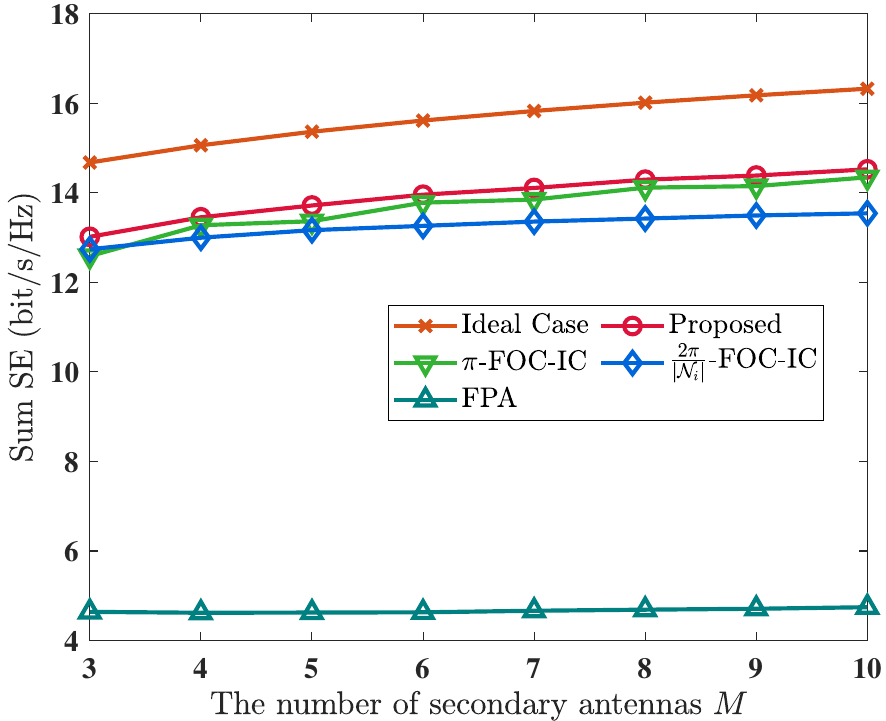}
				\label{fig6}
			}
			\caption{Sum SE versus (a) the number of primary PAs and (b) the number of secondary PAs.}
			\label{Fig-All}
		\end{figure*}			 
		
		\subsection{Impact of the Number of Primary and Secondary PAs}	
		Figure~\ref{Fig-All} illustrates the sum SE performance of the considered algorithms versus the number of primary and secondary antennas, respectively. Both the \textbf{Proposed} and \textbf{FPA} schemes exhibit monotonic improvement with increasing antennas, while the \textbf{Proposed} scheme consistently achieves higher SEs, particularly for larger PA/antenna sizes. The performance of the \textbf{FPA} scheme is limited by the fixed antenna positions, which prevent reductions in large-scale path loss, whereas the proposed PA-enabled CR network effectively exploits additional spatial DoF for signal enhancement, yielding significant system-level gains.
		
		The \textbf{$\pi$-FOC-IC} algorithm exhibits a non-monotonic trend. The performance improves when the number of PAs increases from 3 to 4 but degrades as it grows from 4 to 5. This degradation stems from residual interference associated with a single PA under odd-numbered configurations, which cannot be fully canceled. In contrast, the \textbf{Proposed} algorithm achieves consistent performance gains across all PA settings, confirming its superiority over the \textbf{$\pi$-FOC-IC} approach.

        For small numbers of PAs, the \textbf{Proposed} and \textbf{$\tfrac{2\pi}{|\mathcal{N}_i|}$-FOC-IC} algorithms deliver comparable performance. However, the performance gap widens as the number of PAs increases. When ${|\mathcal{N}_i|}$ is small, the adjacent phase increment $\Delta{\phi}=2\pi/{|\mathcal{N}_i|}$ exceeds the phase error $\Delta \varepsilon$ (i.e., $\Delta {\phi} > \Delta \varepsilon$), so the induced perturbation is minor and interference cancellation is nearly complete. With larger ${|\mathcal{N}_i|}$, $\Delta\phi$ decreases and becomes comparable to or smaller than $|\varepsilon|$ (i.e., $\Delta \phi << \Delta \varepsilon$), which amplifies the perturbation, renders cancellation incomplete, and degrades the achievable sum SE. Consequently, the \textbf{Proposed} scheme exhibits superior scalability and reliability as ${| \mathcal{N}_i |}$ grows.     

\begin{figure}[htpb]
	\centering
	\includegraphics[width=3.25in]{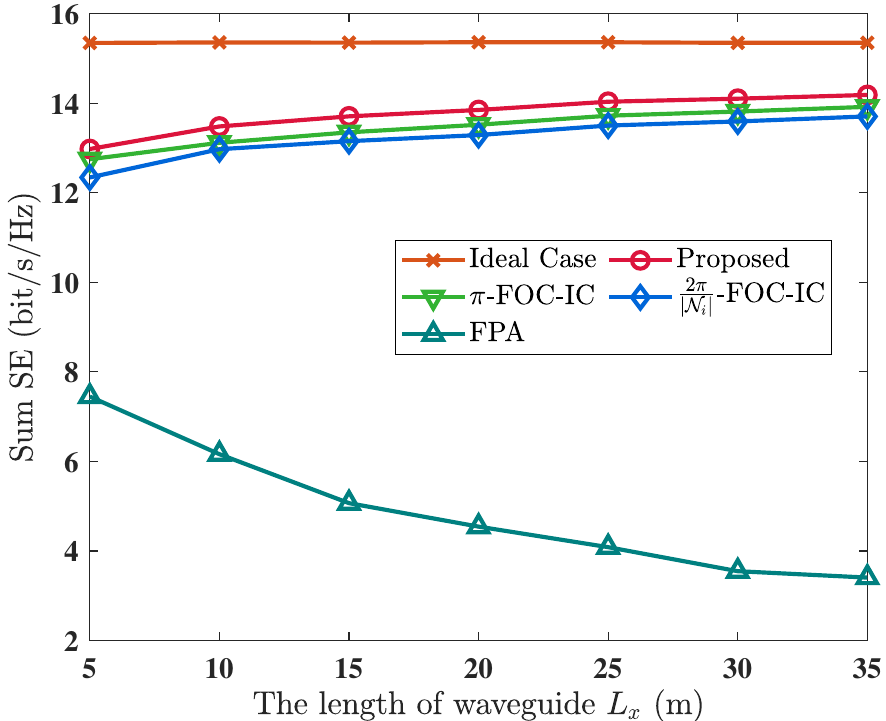}
	\caption{Sum SE versus the length of waveguide $L_x$.}
	\label{fig7}
\end{figure} 	 
		\subsection{Impact of the Waveguide Length }
		Figure~\ref{fig7} shows the impact of the waveguide length on the sum SE. As illustrated in Fig.~\ref{fig7}, the performance of the \textbf{Proposed} algorithm initially improves and then converges to a maximum value. This trend is primarily due to the fact that a longer waveguide leads to a greater average separation between the PU and SU. Consequently, the primary and secondary PAs tend to be positioned closer to their respective receivers, thereby reducing mutual interference. Additionally, the \textbf{Proposed} algorithm achieves performance comparable to that of the \textbf{Ideal Case} benchmark, indicating that the \textbf{Proposed} algorithm can attain a near-ideal performance. The sum SE performance of the \textbf{FPA} scheme exhibits a downward trend due to the increased $L_x$ caused by the higher path loss.  
		
		\subsection{Impact of the Power Budget of PT and ST}
\begin{figure*}[htbp]
			\centering
			\subfigure[]{
				\includegraphics[width=0.46\linewidth]{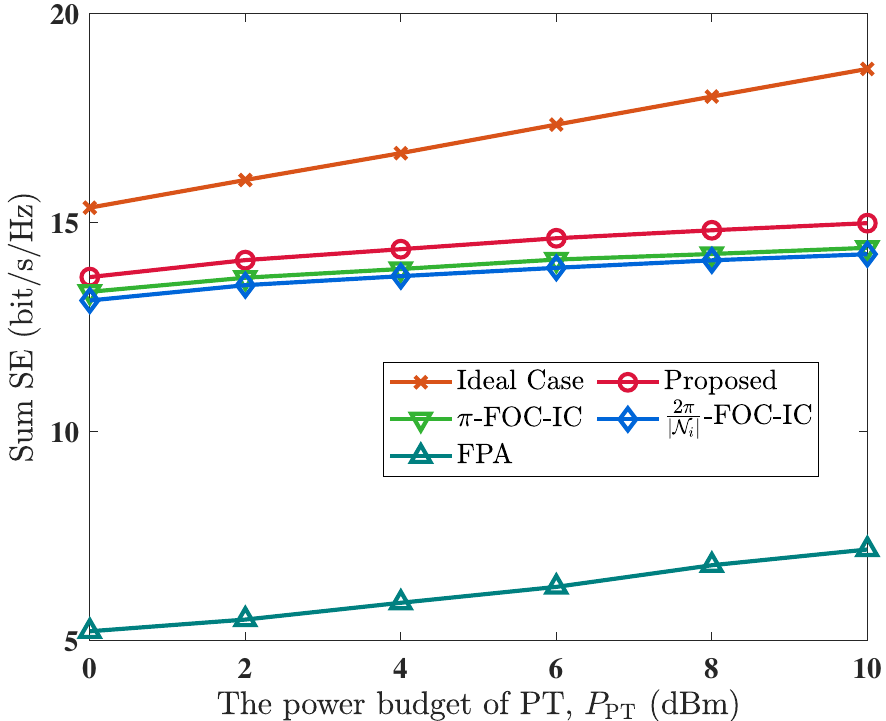}
				\label{fig8}
			}
			\hspace{0.03\linewidth}
			\subfigure[]{
				\includegraphics[width=0.46\linewidth]{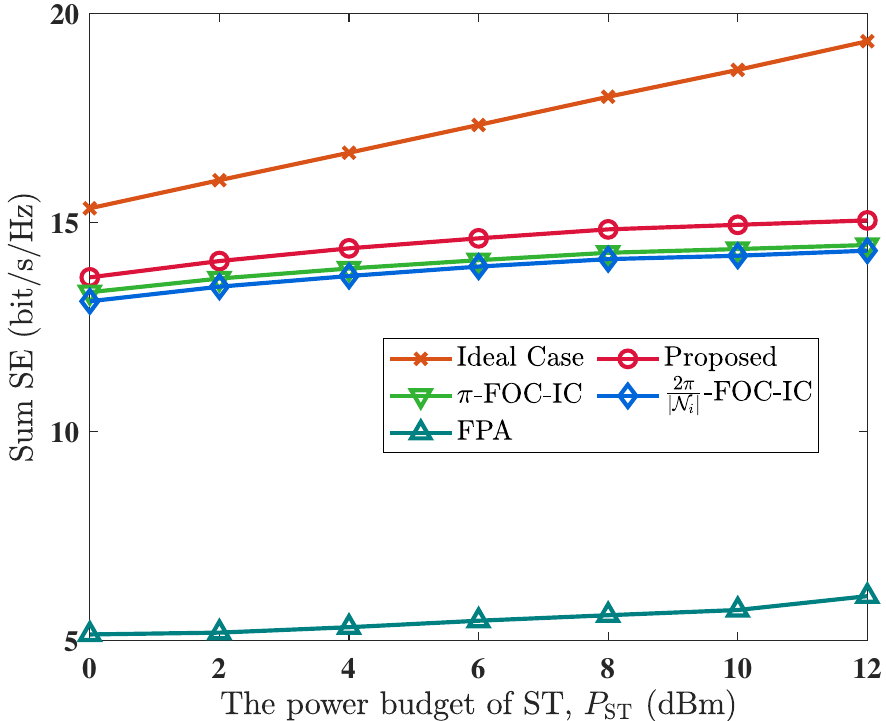}
				\label{fig9}
			}
			\caption{Sum SE versus (a) the power budget of PT and (b) the power budget of ST.} 
			\label{Fig-ab}
		\end{figure*}

		Figure~\ref{Fig-ab} depicts the relationship between the sum SE with the power budget of PT and ST, respectively. As shown in Fig.~\ref{Fig-ab}, all schemes improve monotonically as the PT/ST power increases. Moreover, the \textbf{Proposed} algorithm outperforms \textbf{$\theta_{\mathrm{fix}}$-FOC-IC} algorithms which demonstrates the effectiveness and superiority of the proposed interference cancellation scheme. The performance disparity between the \textbf{Proposed} algorithm and \textbf{Ideal Case} due to the amplified power budget will cause severe interference. The \textbf{Proposed} algorithm consistently surpasses the \textbf{FPA} scheme, confirming the effectiveness of the PA-enabled CR architecture and the proposed interference-cancellation design.

		\section{Conclusion}		
		In this paper, we investigated a PA-enabled CR network, in which both PT and ST are equipped with single waveguide and multiple PAs to facilitate simultaneous spectrum sharing. A joint optimization problem was formulated to maximize the sum ASE of the PU and SU, subject to power, spacing, and interference temperature constraints. To tackle the resulting non-convex problem, we developed a three-stage optimization framework that sequentially determined the pinching beamforming of the PT and ST, followed by ST power control. The proposed design integrated waveguide-level placement with wavelength-level phase alignment, and further incorporated a refined phase assignment strategy to mitigate residual interference in odd-sized PA arrays. Simulation results demonstrated that: i) the proposed PA-enabled CR network yields significant performance gains compared with fixed-position antenna systems; ii) the proposed signal management strategy is particularly effective under odd-sized or small PA configurations; and iii) as the separation between PN and SN increases, the achieved performance closely approaches the theoretical upper bound, thereby enabling near-orthogonal transmissions between primary and secondary links. These findings highlighted the potential of PA systems as a promising paradigm for future spectrum sharing networks.
		
		\appendices
		\section{Proof of Formulas~(\ref{eq7_2}) and~(\ref{eq7_3})}  
		Based on the adopted Ricean fading model and \textit{Lemma} 1, for arbitrary PA $t \in {\cal N}_i$ and user $r \in \mathcal R$, the expectation expression $\mathbb E \left\{ {{{\left| {\sum\limits_t {{h_{t,r}}{g_t}} } \right|}^2}} \right\}$ can be equivalently expressed as
		\begin{equation}
			\begin{aligned}
				\mathbb{E}\left\{ {{{\left| {\sum\limits_t {{h_{t,r}}{g_t}} } \right|}^2}} \right\}\mathop {{\rm{  }} = }\limits^{(a)} \frac{\eta }{{\kappa  + 1}}\left[ {\kappa {{\left| {\sum\limits_t {\frac{{{{\bar h}_{t,r}}{g_t}}}{{{{\left\| {{{\bf{u}}_r} - {{\bf{p}}_t}} \right\|}^{\frac{\chi }{2}}}}}} } \right|}^2}} \right.\\
				\left. { + \sum\limits_t {\frac{1}{{{{\left\| {{{\bf{u}}_r} - {{\bf{p}}_t}} \right\|}^\chi }}}} } \right],
			\end{aligned}
			\label{eq100}
		\end{equation}
		where $(a)$ follows since $\{\tilde h_{t,r}\}$ are i.i.d.\ $\mathcal{CN}(0,1)$, mutually independent across $t$ and independent of $\{\bar h_{t,r}\}$. Therefore, $\mathbb{E}\!\left[\bar h_{t,r}g_t (\tilde h_{t',r}g_{t'})^{\!*}\right]\!=\!0$ for all $t,t'$ and $\mathbb{E}\!\left[\tilde h_{t,r}g_t (\tilde h_{t,r}g_{t'})^{\!*}\right]\!=\!0$ for $t\!\neq\! t'$, leaving the squared magnitude of the coherent LoS superposition and the aggregate NLoS power.

		\section{Proof of Formula~(\ref{eq16_e}) } 
		Consider $N$ complex phasors of unit amplitude,
		\begin{equation}
			z_n = e^{j\left(\phi_0 + n\theta\right)}, \quad n = 1, \ldots, N.
		\end{equation}
		
		The sum of these phasors is given by
		\begin{equation}
			S = \sum_{n=1}^{N} z_n = \sum_{n=1}^{N} e^{j\left(\phi_0 + n\theta\right)} = e^{j\phi_0} \sum_{n=1}^{N} e^{j n \theta}=  e^{j\phi_0} \frac{1 - e^{jN\theta}}{1 - e^{j\theta}}.
		\end{equation}
		
		Recognizing the inner sum as a finite geometric series, we have
		\begin{equation}
			S =  e^{j\phi_0} \sum_{n=1}^{N} e^{j n \theta} =  e^{j\phi_0} \frac{1 - e^{jN\theta}}{1 - e^{j\theta}}.
		\end{equation}
		
		The magnitude of this sum is
		\begin{equation}
			|S| = \left| \frac{1 - e^{jN\theta}}{1 - e^{j\theta}} \right| = \left| \frac{\sin\left(\frac{N\theta}{2}\right)}{\sin\left(\frac{\theta}{2}\right)} \right|.
		\end{equation}
		
		The minimum value of $|S|$ is achieved when the numerator vanishes, i.e., $\sin\left(\frac{N\theta}{2}\right) = 0$, which occurs at
		\begin{equation}
			\theta = \frac{2\pi k}{N}, \quad k = 1, 2, \ldots, N.
		\end{equation}
		
		At these points, the phases of the $N$ elements are uniformly distributed on the unit circle, and thus
		\begin{equation}
			\sum_{n=0}^{N-1} e^{j\left(\phi_0 + n\frac{2\pi}{N}\right)} = 0.
		\end{equation}

		\bibliographystyle{IEEEtran}
		\bibliography{Ref}
		

	\end{document}